\documentclass[preprintnumbers, floatfix,twocolumn,preprintnumbers, letterpaper, superscriptaddress,nofootinbib]{revtex4}
\usepackage{eurosym}
\usepackage{amsfonts}
\usepackage{amsmath}
\usepackage{amssymb,epsf}
\usepackage{latexsym}
\usepackage{graphicx,epsfig}
\usepackage{amssymb}
\usepackage{float}
\usepackage{subfigure}
\usepackage{epstopdf}
\usepackage[colorlinks=true,citecolor=blue,linkcolor=blue,urlcolor=black]{hyperref}
\usepackage{dcolumn}
\usepackage{psfrag}
\usepackage{wrapfig}
\usepackage{makeidx}
\usepackage{epsf}

\begin{document}

\title{One-dimensional backreacting holographic superconductors \\
with exponential nonlinear electrodynamics}

\author{B. Binaei Ghotbabadi}
\affiliation{Physics Department and Biruni Observatory, College of Sciences, Shiraz
University, Shiraz 71454, Iran}

\author{M. Kord Zangeneh}
\email{mkzangeneh@scu.ac.ir}
\affiliation{Physics Department, Faculty of Science, Shahid Chamran University of Ahvaz,
Ahvaz 61357-43135, Iran}
\affiliation{Research Institute for Astronomy and Astrophysics of Maragha (RIAAM),
Maragha, P. O. Box: 55134-441, Iran}

\author{A. Sheykhi}
\email{asheykhi@shirazu.ac.ir}
\affiliation{Physics Department and Biruni Observatory, College of Sciences, Shiraz
University, Shiraz 71454, Iran}
\affiliation{Research Institute for Astronomy and Astrophysics of Maragha (RIAAM),
Maragha, P. O. Box: 55134-441, Iran}

\begin{abstract}
In this paper, we investigate the effects of nonlinear exponential
electrodynamics as well as backreaction on the properties of one-dimensional 
$s$-wave holographic superconductors. We continue our study both
analytically and numerically. In analytical study, we employ the
Sturm-Liouville method while in numerical approach we perform the shooting
method. We obtain a relation between the critical temperature and chemical
potential analytically. Our results show a good agreement between analytical
and numerical methods. We observe that the increase in the strength of both
nonlinearity and backreaction parameters causes the formation of
condensation in the black hole background harder and critical temperature
lower. These results are consistent with those obtained for two dimensional $%
s$-wave holographic superconductors.
\end{abstract}

\maketitle

\section{Introduction}

The $AdS/CFT$ duality provides a correspondence between a strongly coupled
conformal field theory (CFT) in $d$-dimensions and a weakly coupled gravity
theory in ($d+1$)-dimensional anti-de Sitter ($AdS$) spacetime \cite{1,2,3}.
Since it is a duality between two theories with different dimensions, it is
commonly called holography. The idea of holography has been employed in
condensed matter physics to study various phenomena such as superconductors~%
\cite{4,5,6,7,8}. For describing the properties of low temperature
superconductors, the BCS theory can work very well \cite{9,10}. However,
this theory fails to describe the mechanism of high temperature
superconductor. In latter regime, the holography was suggested to study the
properties of superconductors \cite{11,12}. Hortnol $et$ $al.$ have
represented the first model of holographic superconductors~\cite{11,12}.
After that, holographic superconductors have attracted a lot of attention
and investigated from different point of views~\cite{4,13,14,15,16}.

BTZ black holes play a significant role in many of recent developments in
string theory~\cite{46,47,48}. BTZ-like solutions are dual of ($1+1$%
)-dimensional holographic systems such as one-dimensional holographic
superconductors. Distinctive features of normal and superconducting phases
of one-dimensional systems have been studied studied in~\cite{49}. The
latter study was done in probe limit. Considering the effects of
backreaction, the properties of one-dimensional holographic superconductors
have been studied both~numerically \cite{50,50-1} and analytically~\cite%
{51,52}.

It is interesting to investigate the effect of nonlinear electrodynamic
models on holographic systems including holographic superconductors \cite%
{17,18,19,20,21,22,23,24,34,30,31,32,38,Doa,n3,n5,n1,n2,n4,SheyPM}.
Nonlinear models carry more information than the usual Maxwell case and also
are considered as a possible way for avoiding the singularity of the
point-like charged particle at the origin \cite{25,26,27,28,29}. The oldest
nonlinear electrodynamic model is Born-Infeld (BI) model. There are also two
BI-like nonlinear electrodynamics namely logarithmic~\cite{33,nn,34} and
exponential~\cite{35,36,n5} electrodynamics. It has been found that the
exponential electrodynamics has stronger effect on the condensation~than
other models \cite{37}.

In the present work, we will study the one-dimensional holographic
superconductors both analytically and numerically in the presence of
exponential electrodynamics. To bring rich physics in holographic model, we
consider the backreaction of the scalar and gauge fields on the metric
background \cite{39,40,41,42,43,44,45}. To perform the analytical study, we
employ the Sturm-Liouville eigenvalue problem. We will study the effects of
nonlinear exponential electrodynamics model as well as backreaction on
critical temperature. We shall also use the numerical shooting method to
investigate the features of our holographic superconductors and make
comparison between analytical and numerical results.

This paper is organized as follows: In next section, we introduce the action
and basic field equations governing ($1+1$)-dimensional holographic
superconductors in the presence of exponential electrodynamics. In section %
\ref{analst}, we study the properties of holographic superconductors
applying the analytical method based on Sturm-Liuoville eigenvalue problem.
In section \ref{numst}, we study holographic superconductors numerically by
employing the shooting method. We also compare our numerical and analytical
results. Finally, in last section we will summarize our results.

\section{Holographic Set-up}

To study a ($1+1$)-dimensional holographic superconductor, we consider a ($%
2+1$)-dimensional bulk action of AdS gravity coupled to a charged scalar
field $\psi $%
\begin{eqnarray}
S &=&\frac{1}{2\kappa ^{2}}\int d^{3}x\sqrt{-g}\left(R+\frac{2}{l^{2}}\right)
\notag \\
&&+\int d^{3}x\sqrt{-g}\left[ L\left( F\right) -|\nabla \psi -iqA\psi
|^{2}-m^{2}|\psi |^{2}\right] ,  \notag \\
&&  \label{Act}
\end{eqnarray}%
where $g$ is the determinant of metric, $R$ is Ricci scalar, $l$ is the AdS
radius, $A$ is electromagnetic potential and $F=F_{\mu \nu }F^{\mu \nu }$ in
which $F_{\mu \nu }=\nabla _{\lbrack \mu }A_{\mu ]}$. In action (\ref{Act}), 
$\kappa ^{2}=8\pi G_{3}$ where $G_{3}$ is the ($2+1$)-dimensional Newtonian
constant and $m$ and $q$ represent the mass and charge of scalar field,
respectively. $L\left( F\right) $ stands for the Lagrangian of
electrodynamics model. ($1+1$)-dimensional holographic superconductors in
the presence of linear Maxwell electrodynamics presented by $L\left(
F\right) =-F/4$ have been studied in \cite{50, 53}. The linear model is an
idealization of reality. In principle, other powers of $F$ may play role.
There are different nonlinear electrodynamics models which exhibit the
electrodynamics interaction. In this paper, we suppose that electrodynamics
interaction is governed by exponential nonlinear electrodynamics model \cite%
{35}%
\begin{equation}
L\left( F\right) =\frac{1}{4b}\left( {\mathrm{e}}^{-bF}-1\right),
\label{eleclag}
\end{equation}%
where $b$ determines the nonlinearity. For small values of $b$, Lagrangian (%
\ref{eleclag}) recovers the linear Maxwell Lagrangian. The parameter $\kappa 
$ in (\ref{Act}) also determines the backreaction. When $\kappa $ goes to
zero, we are in the probe limit, meaning that the gravity part of action (%
\ref{Act}) is stronger than the matter field part. Physically, this implies
that the gauge and matter fields do not back react on the metric background.
In superconducting language, implies that the Cooper pairs have negligible
interaction with background system. In the presence of backreaction, the
dual black hole solution may be given by the ansatz%
\begin{equation}
ds^{2}=-f(r){\mathrm{e}}^{-\chi (r)}dt^{2}+\frac{dr^{2}}{f(r)}+\frac{r^{2}}{%
l^{2}}dx^{2}.
\end{equation}%
The Hawking temperature of above black hole solution is given by%
\begin{equation}
T=\frac{f^{\prime }(r_{+}){\mathrm{e}}^{-\chi (r_{+})}}{4\pi },  \label{3}
\end{equation}%
where $r_{+}$ is event horizon which could be obtained as the greatest root
of $f(r)=0$. Varying the action (\ref{Act}) with respect to $\psi $, $A_{\nu
}$ and $g_{\mu \nu }$, the field equations read, respectively, 
\begin{eqnarray}
0 &=&\left( \nabla _{\mu }-i{q}A_{\mu }\right) \left( \nabla ^{\mu }-i{q}%
A^{\mu }\right) \psi -m^{2}\psi \,,  \label{01} \\
&&  \notag \\
0 &=&\nabla ^{\mu }\left( 4L_{F}F_{\mu \nu }\right)  \notag \\
&&-i{q}\left[ -\psi ^{\ast }(\nabla _{\nu }-i{q}A_{\nu })\psi +\psi (\nabla
_{\nu }+i{q}A_{\nu })\psi ^{\ast }\right] \,,  \label{02} \\
&&  \notag \\
0 &=&\frac{1}{2\kappa ^{2}}\left[ R_{\mu \nu }-g_{\mu \nu }\left( \frac{R}{2}%
+\frac{1}{l^{2}}\right) \right] +2F_{ac}F_{b}{}^{c}L_{F}  \notag \\
&&-\frac{g_{\mu \nu }}{2}\left[ L\left( F\right) -m^{2}|\psi |^{2}-{|\nabla
\psi -i{q}A\psi |^{2}}\right]  \notag \\
&&-\frac{1}{2}\left[ (\nabla _{\mu }\psi -i{q}A_{\mu }\psi )(\nabla _{\nu
}\psi ^{\ast }+i{q}A_{\nu }\psi ^{\ast })+\mu \leftrightarrow \nu \right] , 
\notag \\
&&  \label{03}
\end{eqnarray}%
where $L_{F}=\partial L/\partial F$. Adopting the ansatz $A_{\mu }=\phi
(r)\delta _{\mu }^{0}$ and $\psi =\psi (r)$, field equations (\ref{01})-(\ref%
{03}) lead to

\begin{eqnarray}
0 &=&\psi ^{\prime \prime }+\psi ^{\prime }\left[ \frac{1}{r}+\frac{%
f^{\prime }}{f}-\frac{\chi ^{\prime }}{2}\right] +\psi \left[ \frac{%
q^{2}\phi ^{2}{\mathrm{e}}^{\chi }}{f^{2}}-\frac{m^{2}}{f}\right] ,
\label{f1} \\
&&  \notag \\
0 &=&\phi ^{\prime \prime }+\phi ^{\prime }\left( \frac{1}{r}+\frac{\chi
^{\prime }}{2}\right)   \notag \\
&&-{\frac{2{q}^{2}\psi ^{2}\phi }{{\mathrm{e}^{2b\phi ^{\prime 2}{\mathrm{e}%
^{\chi }}}}f}+2b\phi ^{\prime 2}{\mathrm{e}^{\chi }}\left( 2\phi ^{\prime
\prime }+\phi ^{\prime }\chi ^{\prime }\right) },  \label{f2} \\
&&  \notag \\
0 &=&f^{\prime }+{\frac{\kappa ^{2}r}{2b}}\left[ 1+{\mathrm{e}^{2b\phi
^{\prime 2}{\mathrm{e}^{\chi }}}}\left( 4b\phi ^{\prime 2}{\mathrm{e}^{\chi }%
}-1\right) \right]   \notag \\
&&+2\,\kappa ^{2}r\left[ {\frac{q^{2}\phi ^{2}\psi ^{2}{\mathrm{e}^{\chi }}}{%
f}}+\left( m\psi \right) ^{2}+\psi ^{\prime 2}f\right] -{\frac{2r}{{l}^{2}}},
\label{f3} \\
&&  \notag \\
0 &=&\chi ^{\prime }+4\kappa ^{2}r\left[ \frac{q^{2}\phi ^{2}\psi ^{2}{%
\mathrm{e}}^{\chi }}{f^{2}}+\psi ^{\prime 2}\right] .  \label{f4}
\end{eqnarray}%
where the prime denotes the derivative with respect to $r$. Obviously, the
above equations reduce to the corresponding equations in Ref.~\cite{50} when 
$b\rightarrow 0$ while in the absence of the backreaction ($\kappa
\rightarrow 0$), Eqs. (\ref{12}) and (\ref{13}) reduce to ones in Ref.~\cite%
{37}. By virtue of symmetries of field equations (\ref{12})-(\ref{15})%
\begin{gather}
q\rightarrow q/a,\text{ \ \ \ \ }\phi \rightarrow a\phi ,\text{ \ \ \ \ }%
\psi \rightarrow a\psi ,  \notag \\
\kappa \rightarrow \kappa /a,\text{ \ \ \ \ }b\rightarrow b/a^{2}, \\
\notag \\
l\rightarrow al,\text{ \ \ \ \ }r\rightarrow ar,\text{ \ \ \ \ }q\rightarrow
q/a,  \notag \\
m\rightarrow m/a,\text{ \ \ \ \ }b\rightarrow a^{2}b,
\end{gather}%
one can set $q=l=1$. In the following sections, we will study the
superconding phase transition both analytically and numerically.

\section{Analytical Study\label{analst}}

The behaviors of model functions governed by field equations (\ref{12})-(\ref%
{15}) near the boundary $r\rightarrow \infty $ are \footnote{%
Near the boundary, $\chi $ could be a constant but by using the symmetry of
field equation $\mathrm{e}^{\chi }\rightarrow a^{2}\mathrm{e}^{\chi },$ $%
\phi \rightarrow \phi /a$, one can set it to zero there.}%
\begin{gather}
\chi (r)\rightarrow 0,\qquad f(r)\sim r^{2},  \notag \\
\phi (r)\sim \rho +\mu \ln (r),\qquad \psi (r)\sim \frac{\psi _{-}}{%
r^{\Delta _{-}}}+\frac{\psi _{+}}{r^{\Delta _{+}}},  \label{boundval}
\end{gather}%
where $\mu $ and $\rho $ are chemical potential and charge density of dual
field theory and $\Delta _{\pm }=1+\sqrt{1\pm m^{2}}$. The superconducting
phase transition is characterized by growing the expectation value of order
parameter $\left\langle O\right\rangle $ as temperature decreases. In normal
phase, $\left\langle O\right\rangle $ vanishes. According to holographic
dictionary, the expectation value of order parameter $\left\langle
O\right\rangle $ is dual to ${\psi _{+}}$ or ${\psi _{-}}$ while the other
one can be considered as the source. Therefore, near the critical point $%
\left\langle O_{\pm }\right\rangle $ is small and one can define it as%
\begin{equation}
\epsilon \equiv \left\langle O_{i}\right\rangle ,
\end{equation}%
where $i=+$ or $-$. Since $\epsilon $ is so small, we can expand the model
functions as~\cite{39,54,55,56}\footnote{%
It is expected that when the sign of $\epsilon $ changes, the sign of scalar
filed which leads to order parameter, changes too. So, the expansion powers
of $\psi $ is considered odd. For other functions, even powers is used
because they should not change when the sign of order parameter changes.} 
\begin{gather}
f=f_{0}+\epsilon ^{2}f_{2}+\epsilon ^{4}f_{4}+\cdots ,  \label{f} \\
\chi =\epsilon ^{2}\chi _{2}+\epsilon ^{4}\chi _{4}+\cdots ,  \label{chi} \\
\psi =\epsilon \psi _{1}+\epsilon ^{3}\psi _{3}+\epsilon ^{5}\psi
_{5}+\cdots ,  \label{psi} \\
\phi =\phi _{0}+\epsilon ^{2}\phi _{2}+\epsilon ^{4}\phi _{4}+\cdots .
\label{phi}
\end{gather}%
Also we can expand the chemical potential as~\cite{55} 
\begin{equation}
\mu =\mu _{0}+\epsilon ^{2}\delta \mu _{2}+...,
\end{equation}%
where $\delta \mu _{2}>0$. Thus, the order parameter as a function of
chemical potential can be obtained as%
\begin{equation}
\epsilon \approx \frac{(\mu -\mu _{0})^{1/2}}{\delta \mu _{2}}.
\end{equation}%
When $\mu \rightarrow \mu _{0}$, phase transition occurs and the order
parameter is zero at the critical value $\mu _{c}=\mu _{0}$. Above equation
also indicates the critical exponent $\beta =1/2$ which is the same as the
universal result from the mean field theory. Hereafter, we define the
dimensionless coordinate $z=r_{+}/r$ instead of $r,$ since it is easier to
work with it. In terms of this new coordinate, $z=0$ and $z=1$ correspond to
the boundary and horizon respectively. The field equations (\ref{12})-(\ref%
{15}) can be rewritten in terms of $z$ as 
\begin{eqnarray}
0 &=&\psi ^{\prime \prime }+\left[ \frac{1}{z}+\frac{f^{\prime }}{f}-\frac{%
\chi ^{\prime }}{2}\right] \psi ^{\prime }+\frac{r_{+}^{2}}{z^{4}}\left[ 
\frac{q^{2}\phi ^{2}{\mathrm{e}}^{\chi }}{f^{2}}-\frac{m^{2}}{f}\right] \psi
,  \notag \\
&&  \label{12} \\
0 &=&\phi ^{\prime \prime }+\left( \frac{\chi ^{\prime }}{2}+\frac{1}{z}%
\left[ \frac{1+4\Upsilon }{1+2\Upsilon }\right] \right) \phi ^{\prime } -%
\frac{2q^{2}r_{+}^{2}\psi ^{2}}{z^{4}f}\left( \frac{{\mathrm{e}}^{-\Upsilon }%
}{1+2\Upsilon }\right) \phi ,  \notag \\
&&  \label{13} \\
0 &=&f^{\prime }-{\frac{\kappa ^{2}{r_{+}}^{2}}{2bz^{3}}}\left[ 1+{\mathrm{e}%
^{\Upsilon }}\left( {2\Upsilon }-1\right) \right]  \notag \\
&&-\frac{2\kappa ^{2}{r_{+}^{2}}}{z^{3}}\left[ \frac{q^{2}\phi ^{2}{\psi ^{2}%
}{\mathrm{e}^{\chi }}}{f}+m^{2}\psi ^{2}\right] +{\frac{2{r_{+}^{2}}}{z^{3}}}%
,  \label{14} \\
&&  \notag \\
0 &=&\chi ^{\prime }-\frac{4\kappa ^{2}r_{+}^{2}}{z^{3}}\left[ \frac{%
q^{2}\phi ^{2}\psi ^{2}{\mathrm{e}}^{\chi }}{f^{2}}+\frac{z^{4}\psi ^{\prime
2}}{r_{+}^{2}}\right] ,  \label{15}
\end{eqnarray}%
where $\Upsilon =2bz^{4}{\mathrm{e}}^{\chi }\phi ^{\prime 2}/{r_{+}}^{2}$.
The field equation of $\phi $ (Eq. (\ref{13})) at zeroth order with respect
to $\epsilon $ reduces to%
\begin{equation}
\phi ^{\prime \prime }(z)+\frac{\phi ^{\prime }(z)(r_{+}^{2}+8{b}{z^{4}}\phi
^{\prime 2}(z))}{z(r_{+}^{2}+4{b}{z^{4}}\phi ^{\prime 2}(z))}=0.
\end{equation}%
The solution of above equation reads%
\begin{equation}
\phi (z)=\int_{1}^{z}\frac{r_{+}\sqrt{L_{W}(\frac{4bz^{2}C_{0}^{2}}{{r_{+}}%
^{2}})}}{2{z^{2}}\sqrt{b}}dz,  \label{24}
\end{equation}%
where $C_{0}$ is an integration constant and $L_{W}(x)=LambertW(x)$ is the
Lambert function which satisfies~\cite{57} 
\begin{equation}
L_{W}(x){\mathrm{e}}^{L_{W}(x)}=x,
\end{equation}%
and can be expanded as%
\begin{equation}
L_{W}(x)=x-x^{2}+\frac{3}{2}x^{3}-\frac{8}{3}x^{4}.
\end{equation}%
Expanding Eq. (\ref{24}) for small $b$ and keeping the terms up to first
order of $b$ we find%
\begin{equation}
\phi (z)=C_{1}+C_{0}\ln (z)+\frac{C_{0}^{3}}{r_{+}^{2}}(1-z^{2})b+O\left(
b^{2}\right) .  \label{27}
\end{equation}%
Comparing the above equation with Eq. (\ref{boundval}), we find $C_{0}=-\mu $%
. Also, $C_{1}=0$, since at the horizon $\phi(r_{+})=0$. Inserting $C_{0}$
into Eq. (\ref{27}) we have 
\begin{equation}
\phi _{0}(z)=\lambda r_{+}\left[ -\ln (z)+\lambda ^{2}(z^{2}-1)b\right],%
\text{ \ \ \ \ }{b}\lambda ^{2}<1,  \label{phi0}
\end{equation}%
where $\lambda =\mu /r_{+}$. Substituting $\phi (z)$ into the field equation
(\ref{14}), we find the metric function at zeroth order with respect to $%
\epsilon $, $f_{0}(z)=r_{+}^{2}g(z)$ where%
\begin{equation}
g(z)=\left( \frac{1}{z^{2}}-1+\kappa ^{2}\lambda ^{2}\ln (z)+\frac{1}{2}%
\kappa ^{2}\lambda ^{4}b(1-z^{2})\right).
\end{equation}%
Note that at the horizon $f_{0}(1)=0$. The asymptotic behavior of scalar
field $\psi $ near the boundary ($z=0$) is given by Eq. (\ref{boundval}). In
order to match this behavior near the boundary, we introduce a trial
function $F(z)$ as $\psi (z)=\left\langle O_{i}\right\rangle \left(
z/r_{+}\right) ^{\Delta _{i}}F(z)$ which satisfies the boundary condition $%
F(0)=1$ and $F^{\prime }(0)=0$. Inserting the functions obtained above and
the trial function $F(z)$ into Eq. (\ref{12}), one receives%
\begin{align}
& F^{\prime \prime }(z)+F^{\prime }(z)\left[ \frac{2\Delta +1}{z}+\frac{%
g^{\prime }(z)}{g(z)}\right]  \notag \\
& +\left[ -\frac{m^{2}}{g(z)z^{4}}+\frac{\Delta ^{2}}{z^{2}}+\frac{\Delta
g^{\prime }(z)}{zg(z)}\right] F(z)+  \notag \\
& +\frac{q^{2}\lambda ^{2}F(z)}{g(z)^{2}}\left[ \frac{{\ln (z)}^{2}}{z^{4}}-%
\frac{2\lambda ^{2}{b}{\ln (z)}}{z^{2}}+\frac{2\lambda ^{2}{b}{\ln (z)}}{%
z^{4}}\right] =0.
\end{align}%
It is a matter of calculations to show that the above equation satisfies the
following second order Sturm-Liouville equation~\cite{58} 
\begin{equation}
\left[ T(z)F^{\prime }(z)\right] ^{\prime }-Q(z)F(z)+\lambda ^{2}P(z)F(z)=0,
\label{35}
\end{equation}%
where 
\begin{eqnarray}
T(z) &=&{g(z)}z^{2\Delta +1}, \\
P(z) &=&\frac{q^{2}T(z)}{g^{2}(z)}\left[ \frac{{\ln (z)}^{2}}{z^{4}}-\frac{%
2\lambda ^{2}{b}{\ln (z)}}{z^{2}}+\frac{2\lambda ^{2}{b}{\ln (z)}}{z^{4}}%
\right] ,  \notag \\
&& \\
Q(z) &=&-T(z)\left[ -\frac{m^{2}}{g(z)z^{4}}+\frac{\Delta ^{2}}{z^{2}}+\frac{%
\Delta g^{\prime }(z)}{zg(z)}\right] .
\end{eqnarray}%
Considering the trial function as $F(z)=1-\alpha z^{2}$ and using the
Sturm-Liouville eigenvalues problem, the eigenvalues of Eq. (\ref{35}) can
be obtained by minimizing%
\begin{equation}
\lambda ^{2}=\frac{\int_{0}^{1}{dz}T(z)\left[ F^{\prime 2}(z)+Q(z)F^{2}(z)%
\right] }{\int_{0}^{1}{dz}.T(z)P(z)F^{2}(z)},  \label{39}
\end{equation}%
with respect to $\alpha $~\cite{59}. Here, we use a perturbative expansion $%
b\lambda ^{2}$ up to the first order of $b$%
\begin{equation}
b\lambda ^{2}=b(\lambda ^{2}|_{b=0})+O(b^{2}).
\end{equation}%
For backreaction parameter, we use iteration method and take%
\begin{equation}
\kappa _{n}=n\Delta \kappa ,\qquad \ n=0,1,2,3,\cdots ,
\end{equation}%
where $\Delta \kappa =\kappa _{n+1}-\kappa _{n}$. Here we chose $\Delta
\kappa =0.005$. So, the effect of the nonlinear corrections on the
backreaction term can be obtained as%
\begin{equation}
\kappa ^{2}\lambda ^{2}={\kappa _{n}}^{2}\lambda ^{2}={\kappa _{n}}%
^{2}(\lambda ^{2}|_{\kappa _{n-1}})+O[(\Delta \kappa )^{4}].
\end{equation}%
By taking $\kappa _{-1}=0$ and $\lambda ^{2}|_{\kappa _{-1}}=0$, the minimum
eigenvalue of Eq. (\ref{39}) can be calculated. To calculate the critical
temperature, we obtain the latter value by variation of Eq. (\ref{39}) with
respect to $\alpha $ where the other parameters such as $b,\kappa
,m,q,\cdots $ are fixed. Using the definition of $T$ (Eq. (\ref{3})), the
critical temperature is given by $T_{c}=f^{\prime }(r_{+})/4\pi $ where%
\footnote{%
Note that at zeroth order with respect to $\epsilon $, $\chi $ is zero. So, $%
{\mathrm{e}}^{\chi }$ in temperature formula disappears.} 
\begin{equation}
f^{\prime }(r_{+})={{2{r_{+}}}}-{\frac{\kappa ^{2}{r_{+}}}{2b}}\left[ 1+{%
\mathrm{e}^{2b\phi _{0}^{\prime 2}(r_{+})}}\left( 4b\phi _{0}^{\prime
2}(r_{+})-1\right) \right] ,
\end{equation}%
and $r_{+}=\mu /\lambda $ so we have 
\begin{equation}
T_{c}=\frac{1}{4\pi }\frac{\mu }{\lambda }\left[ 2-{\kappa _{n}}^{2}(\lambda
^{2}|_{\kappa _{n-1}})+3b{\kappa _{n}}^{2}(\lambda ^{4}|_{\kappa _{n-1},b=0})%
\right] .  \label{45}
\end{equation}%
As an example, for $b=0.01$ and $\kappa =0$ the equation (\ref{39}) reduce
to 
\begin{equation}
\lambda ^{2}=\frac{0.667\alpha ^{2}-1.333\alpha +1}{0.0276-0.0165\alpha
+0.0035\alpha ^{2}},
\end{equation}%
which has a minimum $\lambda _{\min }=21.6337$, with respect to $\alpha $,
at $\alpha =0.8258$ and thus according to Eq. (\ref{45}) we achieve $%
T_{c}=0.03421\mu $. We employ the iteration method to obtain the critical
temperature for different values of $\kappa $ and $b$.

In table \ref{Table21}, we present our results. This table shows that by
increasing nonlinear parameter ($b$), the value of $T_{c}$ decreases. As it
can be understood from this table, for a fixed value of $b$, with increasing
the backreaction parameter $\kappa $, the value of the critical temperature
decreases. We also reproduce the results of the linear Maxwell case without
backreaction (i.e. $b\rightarrow 0$ and $\kappa \rightarrow 0$) presented in 
\cite{51}.

\begin{table}[t]
\begin{center}
\begin{tabular}{|c|c|c|c|c|c|}
\hline
$b$ & $\kappa^{2}=0$ & $\kappa^{2}=0.005$ & $\kappa^{2}=0.01$ & $%
\kappa^{2}=0.015$ & $\kappa^{2}=0.02$ \\ \hline
0 & 0.0429 & 0.0393 & 0.0379 & 0.0369 & 0.0359 \\ \hline
0.01 & 0.0342 & 0.0321 & 0.0311 & 0.0301 & 0.0288 \\ \hline
0.02 & 0.0274 & 0.0231 & 0.0216 & 0.0208 & 0.0199 \\ \hline
\end{tabular}%
\end{center}
\caption{Analytical results of ${T_{c}}/{\protect\mu }$ for different values
of $\protect\kappa $ and $b$.}
\label{Table21}
\end{table}

\begin{table}[t]
\begin{center}
\begin{tabular}{|c|c|c|c|c|c|}
\hline
$b$ & $\kappa^{2}=0$ & $\kappa^{2}=0.05$ & $\kappa^{2}=0.1$ & $%
\kappa^{2}=0.15$ & $\kappa^{2}=0.2$ \\ \hline
0 & 0.046 & 0.0368 & 0.0295 & 0.0236 & 0.0189 \\ \hline
0.01 & 0.0415 & 0.033 & 0.0262 & 0.0209 & 0.0166 \\ \hline
0.04 & 0.0325 & 0.0252 & 0.0197 & 0.0154 & 0.0121 \\ \hline
0.09 & 0.0236 & 0.0176 & 0.0133 & 0.0107 & 0.0078 \\ \hline
\end{tabular}%
\end{center}
\caption{Numerical results of ${T_{c}}/{\protect\mu }$ for different values
of $\protect\kappa $ and $b$.}
\label{Table31}
\end{table}

\section{Numerical Study\label{numst}}

\begin{figure*}[t]
\centering{%
\subfigure[~$b=0$]{
   \label{fig1a}\includegraphics[width=.46\textwidth]{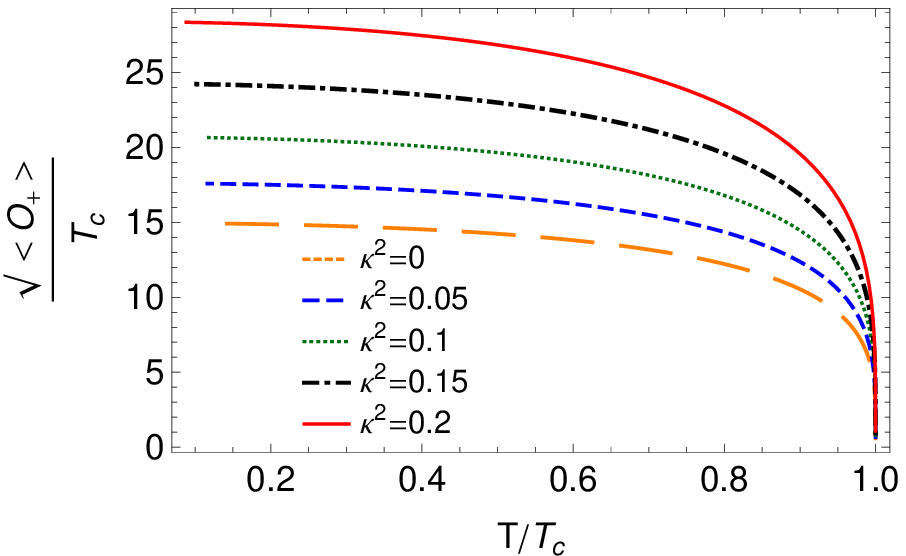}\qquad}}%
\subfigure[~$b=0.04$]{
   \label{fig1b}\includegraphics[width=.46\textwidth]{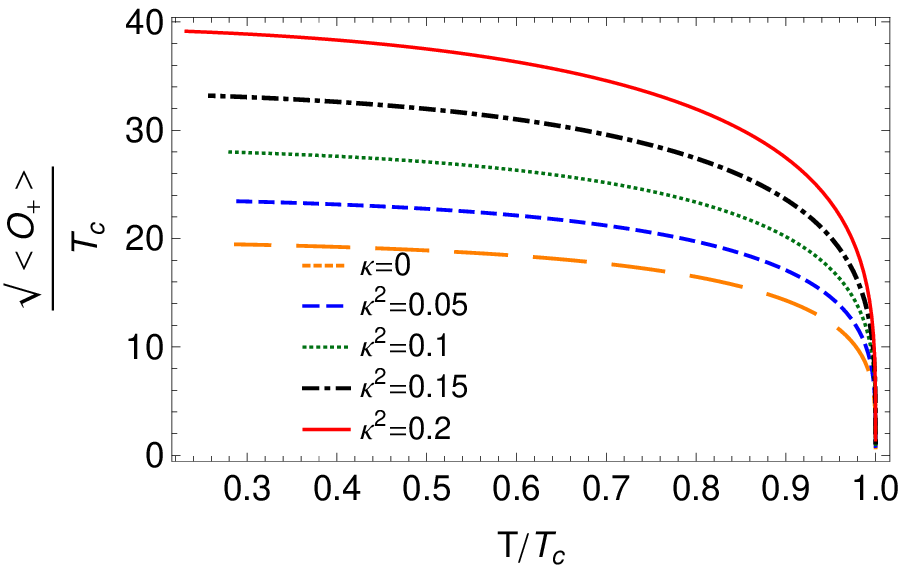}\qquad}
\caption{The behavior of order parameter versus temperature for $m^{2}=0$.}
\label{fig1}
\end{figure*}

\begin{figure*}[t]
\centering{%
\subfigure[~$\protect\kappa ^{2}=0$]{
   \label{fig2a}\includegraphics[width=.46\textwidth]{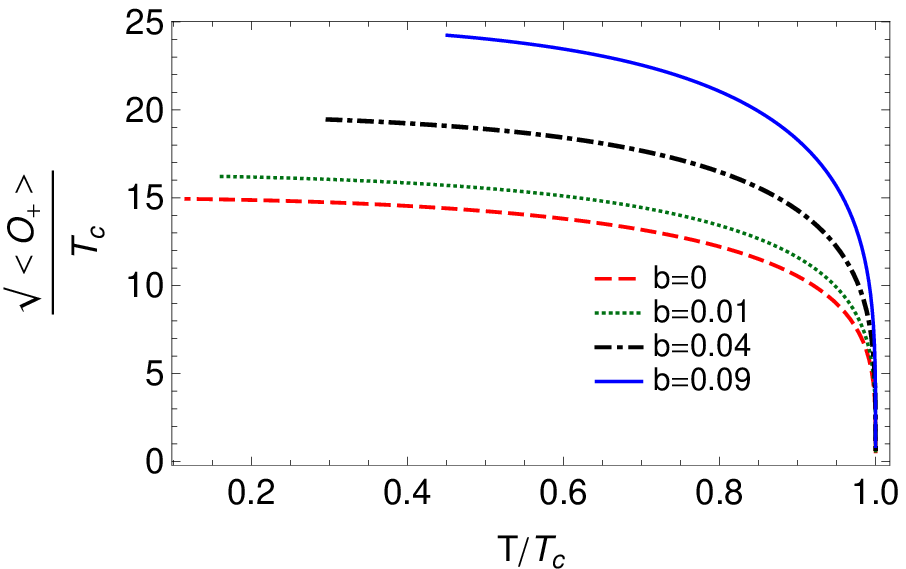}\qquad}}%
\subfigure[~$\protect\kappa ^{2}=0.1$]{
   \label{fig2b}\includegraphics[width=.46\textwidth]{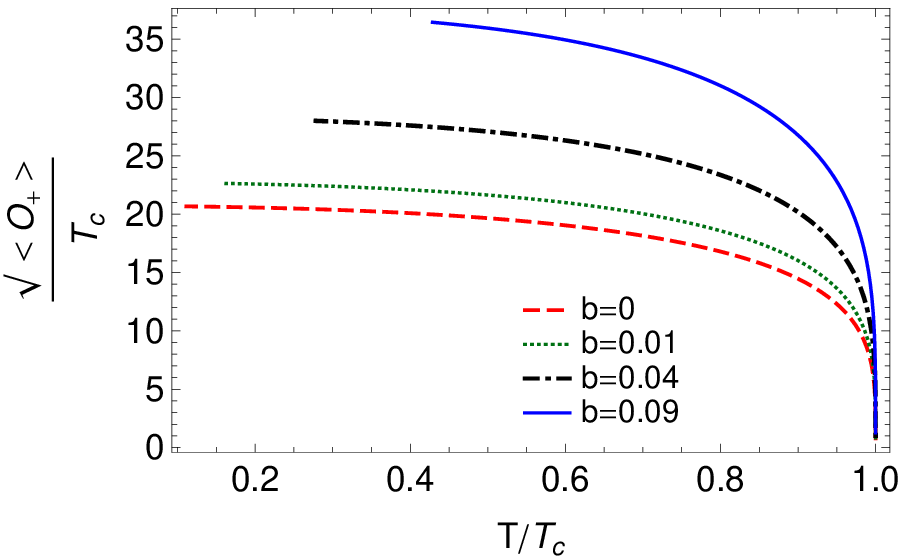}\qquad}
\caption{The behavior of order parameter versus temperature for $m^{2}=0$.}
\label{fig2}
\end{figure*}

\begin{table*}[t]
\centering%
\begin{tabular}{|c|c|c|c|c|c|c|c|c|c|c|}
\hline
& \multicolumn{2}{|c|}{$\kappa ^{2}=0$} & \multicolumn{2}{|c|}{$\kappa
^{2}=0.005$} & \multicolumn{2}{|c|}{$\kappa ^{2}=0.01$} & 
\multicolumn{2}{|c|}{$\kappa ^{2}=0.015$} & \multicolumn{2}{|c|}{$\kappa
^{2}=0.02$} \\ \cline{2-11}\hline
$b$ & $T_{c}$(An) & $T_{c}$(Nu) & $T_{c}$(An) & $T_{c}$(Nu) & $T_{c}$(An) & $%
T_{c}$(Nu) & $T_{c}$(An) & $T_{c}$(Nu) & $T_{c}$(An) & $T_{c}$(Nu) \\ \hline
0 & 0.0429 & 0.046 & 0.0393 & 0.0449 & 0.0379 & 0.0439 & 0.0369 & 0.0430 & 
0.0359 & 0.0409 \\ \hline
0.01 & 0.0342 & 0.0415 & 0.0321 & 0.0406 & 0.0311 & 0.0396 & 0.0301 & 0.0387
& 0.0288 & 0.0378 \\ \hline
0.02 & 0.0274 & 0.0380 & 0.0231 & 0.0371 & 0.0216 & 0.0362 & 0.0208 & 0.0353
& 0.0199 & 0.0345 \\ \hline
\end{tabular}
\caption{Analytical and numerical results of ${T_{c}}/{\protect\mu }$ for
different values of $\protect\kappa $ and $b$.}
\label{Table1}
\end{table*}

In this section, we employ the shooting method \cite{4} to numerically
investigate the superconducting phase transition. Besides setting $q$ and $l$
to unity, we also set $r_{+}=1$ in the numerical calculation which may be
justified by virtue of the field equation symmetry \cite{50}%
\begin{equation*}
r\rightarrow ar,\text{ \ \ \ \ }f\rightarrow a^{2}f,\text{ \ \ \ \ }\phi
\rightarrow a\phi ,
\end{equation*}%
First, we expand Eqs. (\ref{12})-(\ref{15}) near black hole horizon ($z=1$)%
\begin{eqnarray}
\psi  &\approx &\psi (1)+\psi ^{\prime }(1)(1-z)+\frac{\psi ^{{\prime }{%
\prime }}}{2}(1-z)^{2}+\cdots , \\
\phi  &\approx &\phi ^{\prime }(1)(1-z)+\frac{\phi ^{{\prime }{\prime }}}{2}%
(1-z)^{2}+\cdots , \\
f &\approx &f^{\prime }(1)(1-z)+\frac{f^{{\prime }{\prime }}}{2}%
(1-z)^{2}+\cdots , \\
\chi  &\approx &\chi (1)+\chi ^{\prime }(1)(1-z)+\frac{\chi ^{{\prime }{%
\prime }}}{2}(1-z)^{2}+\cdots .
\end{eqnarray}%
In above equations, we have imposed $f(1)=\phi (1)=0$.\footnote{$\phi $
should vanish at horizon so that the norm of gauge potential is regular
there.} In our numerical process, we will find $\psi (1)$, $\phi ^{\prime
}(1)$ and $\chi (1)$ such that the desired values for boundary parameters in
Eq. (\ref{boundval}) are attained. At boundary, one can set either $\psi _{-}
$ or $\psi _{+}$ to zero as source and find the value of the other one as
the expectation value of order parameter $\left\langle O\right\rangle $. We
will focus on $m^{2}=0$ case for our numerical calculations. For this case,
the behavior of $\psi $ near boundary is (see Eq. (\ref{boundval}))%
\begin{equation}
\psi (z)\approx \psi _{-}+{\psi _{+}}{z^{2}}.
\end{equation}%
We consider $\psi _{+}$ as holographic dual to the order parameter $%
\left\langle O_{+}\right\rangle $ at the boundary field theory.

In table \ref{Table31}, our numerical results for critical temperature with
different values of backreaction parameter $\kappa $ and nonlinear parameter 
$b$ are presented. In the Maxwell limit ($b\rightarrow 0)$, our numerical
results reproduce the ones of \cite{50}. It can be seen that in the absence
of nonlinearity effect, the critical temperature decreases as $\kappa $
increases~\cite{50}. In the presence of nonlinearity parameter i.e. for any
nonvanishing value of $b$, it can be found that as $\kappa $ enhances, the
critical temperature $T_{c}$ decreases. Similar behavior can be found for
different values of $b$ when $\kappa $ is fixed. As the nonlinear parameter $%
b$ becomes larger, the critical temperature decreases i.e. the condensation
process becomes harder. This behavior have been reported previously in \cite%
{60} for ($2+1$)-dimensional holographic superconductors too. Figs. \ref%
{fig1} and \ref{fig2} confirm above results. As it can be seen from Fig. \ref%
{fig1}, the scalar hair forms harder as $\kappa $ increases i.e. the gap in
graph of $\left\langle O_{+}\right\rangle $ becomes larger. The latter means
that the condensation of the operator $\left\langle O_{+}\right\rangle $
starts at larger values for stronger values of backreaction parameter. It
shows that the scalar hair can be formed more difficult when the
backreaction is stronger. Fig. \ref{fig2} shows the same effect for
nonlinearity parameter $b$. We compare the analytical and numerical results
in table \ref{Table1}. Table \ref{Table1} shows that there is a good
agreement between analytical and numerical results for small values of $%
\kappa $ and $b$. For larger values of these parameters, analytical and
numerical results separate more from each other.

\section{Conclusion}

In this work, we have studied the properties of one-dimensional holographic
superconductor in the presence of nonlinear exponential electrodynamics. We
have also considered the backreaction effect of scalar and gauge fields on
the background metric. We have performed both analytical and numerical
methods for studying our superconductors. To investigate the problem
analytically, we have used the Sturm-Lioville while our numerical study was
based on shooting method. It was shown that the enhancement in both
nonlinearity of electrodynamics model as well as the backreaction causes the
superconducting phase more difficult to be appeared. This result is
reflected in two ways from our data. From one side, we observed that the
increasement in the effects of nonlinearity and backreaction makes the
critical temperature of superconductor lower. From another side, for larger
values of nonlinear and backreaction parameters, the gap in condensation
parameter is larger which in turn exhibits that the condensation is formed
harder. We have also observed that, for small values of backreaction
parameter $\kappa $ and nonlinear parameter $b$, the analytic results are in
a good agreement with numerical ones whereas for larger values they separate
more from each other.

Finally, we would like to stress that in this work, we have only studied the
basic properties of one-dimensional backreacting holographic $s$-wave
superconductors in the presence of exponential nonlinear electrodynamics. It
is also interesting to investigate other characteristics of these systems
such as the behaviour of real and imaginary parts of conductivity or optical
features. One may also consider $(1+1)$-dimensional $p$-wave and $d$-wave
holographic superconductors in the background of BTZ black holes and
disclose the effects of nonlinearity as well as backreaction on the the
phase transition and conductivity of these models. These issue are now under
investigations and the result will be appeared soon.

\begin{acknowledgments}
MKZ would like to thank Shahid Chamran University of Ahvaz, Iran for
supporting this work. AS thanks the research council of Shiraz University.
The work of MKZ has been supported financially by Research Institute for
Astronomy \& Astrophysics of Maragha (RIAAM) under research project No.
1/5237-55.
\end{acknowledgments}

\end{document}